\documentclass[manuscript=communication]{achemso}

\usepackage{pdfpages}
\usepackage{balance}
\usepackage{physics}
\usepackage{siunitx}

\usepackage{chemformula} 
\usepackage[T1]{fontenc} 

\usepackage{times}

\title{Coupling lattice instabilities across the interface in ultrathin oxide heterostructures}

\author{T.C.~van Thiel}
\email{t.c.vanthiel@tudelft.nl}
\affiliation{Kavli Institute of Nanoscience, Delft University of Technology, Lorentzweg 1, 2628 CJ Delft, Netherlands}

\author{J.~Fowlie}
\affiliation{Department of Quantum Matter Physics, University of Geneva, 24 Quai Ernest-Ansermet, 1211 Gen\`eve 4, Switzerland}

\author{C.~Autieri}
\affiliation{International Research Centre MagTop, Institute of 
Physics, Polish Academy of Sciences, Aleja Lotnik\'{o}w 32/46, PL-02668 
Warsaw, Poland}
\altaffiliation{Consiglio Nazionale delle Ricerche, Istituto 
Superconduttori, Materiali Innovativi e Dispositivi (CNR-SPIN), c/o 
Universit{\`a} G. D'Annunzio, I-66100 Chieti, Italy}

\author{N.~Manca}
\affiliation{Kavli Institute of Nanoscience, Delft University of Technology, Lorentzweg 1, 2628 CJ Delft, Netherlands}

\author{M.~\v{S}i\v{s}kins}
\affiliation{Kavli Institute of Nanoscience, Delft University of Technology, Lorentzweg 1, 2628 CJ Delft, Netherlands}

\author{D.~Afanasiev}
\affiliation{Kavli Institute of Nanoscience, Delft University of Technology, Lorentzweg 1, 2628 CJ Delft, Netherlands}

\author{S.~Gariglio}
\affiliation{Department of Quantum Matter Physics, University of Geneva, 24 Quai Ernest-Ansermet, 1211 Gen\`eve 4, Switzerland}

\author{A.~D.~Caviglia}
\affiliation{Kavli Institute of Nanoscience, Delft University of Technology, Lorentzweg 1, 2628 CJ Delft, Netherlands}

\makeatletter
\renewcommand*{\acs@contact@details}{$^\ast$E-mail:  \acs@email@list}
\makeatother

\renewenvironment{abstract}{%
\vspace{2 mm}%
\rmfamily
\noindent
\emph{\textbf{Abstract:}}%
}{\vspace{0.5em}%
}

\makeatletter
\def\@maketitle{%
\pagestyle{acs}%
\ifnum\acs@author@cnt<\z@\relax
\acs@warning{No authors defined: At least one author is required}%
\fi
\newpage
\null
\vspace*{\acs@space@pre@title}%
\begin{center}
\begin{minipage}{\acs@maketitle@width}
\begin{center}
{%

\titlefont
\titlesize
\let\@fnsymbol\acs@author@fnsymbol
\let\footnote\acs@title@footnote
\acs@maketitle@suppinfo \@title
\acs@title@footnote@check
\global\acs@footnote@cnt\c@footnote
\@maketitle@title@hook
\par
}%
\vspace*{\acs@space@post@title}%
{%
\authorsize
\authorfont
\frenchspacing
\acs@author@list
\par
}%
\vspace*{\acs@space@post@author}%
{%
\affilsize
\affilfont
\acs@address@list
\par
}%
\vspace*{\acs@space@post@address}%
{%
\emailsize
\emailfont
\ifacs@email
\expandafter\acs@contact@details
\fi
}%
\vspace*{\acs@space@post@email}%
\end{center}
\end{minipage}
\end{center}%
}

\renewcommand*\affilfont{\itshape \rmfamily}
\renewcommand*\authorfont{\rmfamily}
\renewcommand*\emailfont{\rmfamily}
\renewcommand*\titlefont{\bfseries \rmfamily}

\AtBeginDocument{
\def\bibsection{
\makeatletter
\@startsection
{section}
{1}
{\z@}{\z@}{2.5mm}
{\vspace{1.5em}\normalfont\acksize\bfseries}
{\nobreak\vspace{1.2mm}\noindent\refname}}
}

\begin{document}


\begin{abstract} \textbf{Oxide heterointerfaces constitute a rich platform for realizing novel functionalities in condensed matter. A key aspect is the strong link between structural and electronic properties, which can be modified by interfacing materials with distinct lattice symmetries. Here we determine the effect of the cubic-tetragonal distortion of \ch{SrTiO3} on the electronic properties of thin films of \ch{SrIrO3}, a topological crystalline metal hosting a delicate interplay between spin-orbit coupling and electronic correlations. We demonstrate that below the transition temperature at 105 K, \ch{SrIrO3} orthorhombic domains couple directly to tetragonal domains in \ch{SrTiO3}. This forces the in-phase rotational axis to lie in-plane and creates a binary domain structure in the \ch{SrIrO3} film. The close proximity to the metal--insulator transition in ultrathin \ch{SrIrO3} causes the individual domains to have strongly anisotropic transport properties, driven by a reduction of bandwidth along the in-phase axis. The strong structure-property relationships in perovskites make these compounds particularly suitable for static and dynamic coupling at interfaces, providing a promising route towards realizing novel functionalities in oxide heterostructures.}
\end{abstract}


Engineering matter with tailored properties is one of the main objectives in materials science. Perovskite oxides have been at the center of attention due to the combination of a flexible lattice structure and strong structure-property relationships. At heterointerfaces, structural phases and domain patterns that are not present in bulk can manifest~\cite{rondinelli2012control, kan2015research, liao2016long}. Such artificial phases can have a marked effect on electronic and magnetic properties and have been shown to modify features such as magnetic anisotropy~\cite{liao2016controlled, kan2016tuning}, interfacial ferromagnetism~\cite{grutter2016interfacial, guo2017interface, paul2014exotic} and ferroelectricity~\cite{kim2013interplay}. Recent years have seen an increasing amount of attention focused on the exploration of nanoscale domains, which have emerged as an abundant source of novel physical properties~\cite{seidel2009conduction,kalisky2013locally, honig2013local, chauleau2019electric, seidel2019nanoelectronics}. Control of such domain patterns however, remains an open challenge. A possible way forward is to incorporate materials that undergo structural phase transitions. A canonical example is \ch{SrTiO3}, a widely used material that undergoes a transition from a cubic to a tetragonal phase when lowering the temperature below $105\;\textrm{K}$. At this temperature, \ch{SrTiO3} breaks up into ferroelastic domains in which \ch{TiO6} octahedra rotate about one of three possible directions~\cite{lytle1964x}. When \ch{SrTiO3} is used as a substrate for heteroepitaxial growth, the rotational distortion and resulting domain pattern can interact with the thin film due to octahedral connectivity across the interface~\cite{segal2011dynamic}. In this context, semimetal \ch{SrIrO3} is of particular interest, since dimensionality and octahedral rotations have been shown to be pivotal in the delicate interplay between spin-orbit coupling (SOC) and electronic correlations~\cite{nie2015interplay, groenendijk2017spin, schutz2017dimensionality}. Efforts to study \ch{SrIrO3} have primarily been fueled by theoretical predictions of a Dirac nodal ring, which is at the boundary between multiple topological classes, depending on the type of lattice symmetry-breaking~\cite{carter2012semimetal, chen2015topological, kim2015surface}. In this respect, the interplay between the correlation strength and electronic bandwidth is crucial as it determines the position of the Dirac point with respect to the Fermi level~\cite{chen2015topological, nie2015interplay, fujioka2019strong}. The bandwidth is,~among other things, governed by the \ch{Ir-O-Ir} bond angle, which may be controlled through cation substitution, pressure tuning~\cite{yamada2019large} or heteroepitaxy.\\
Here we demonstrate manipulation of the structural domain pattern of \ch{SrIrO3} thin films, through interaction with the tetragonal distortion in \ch{SrTiO3}. We find that tetragonal domains in the substrate couple directly to orthorhombic domains in the film, forcing a binary domain structure in \ch{SrIrO3}. In ultrathin films, the \textcolor{black}{\ch{SrTiO3}} tetragonal distortion induces a strong anisotropy in the longitudinal resistivity of \textcolor{black}{\ch{SrIrO3}}, manifesting as a metal-to-insulator transition. \textit{Ab-initio} calculations on ultrathin films corroborate the anisotropic character of the domains, revealing a depletion of states at the Fermi level along one lattice axis, while along the other the system remains metallic.


\begin{figure}
\includegraphics[width=\linewidth]{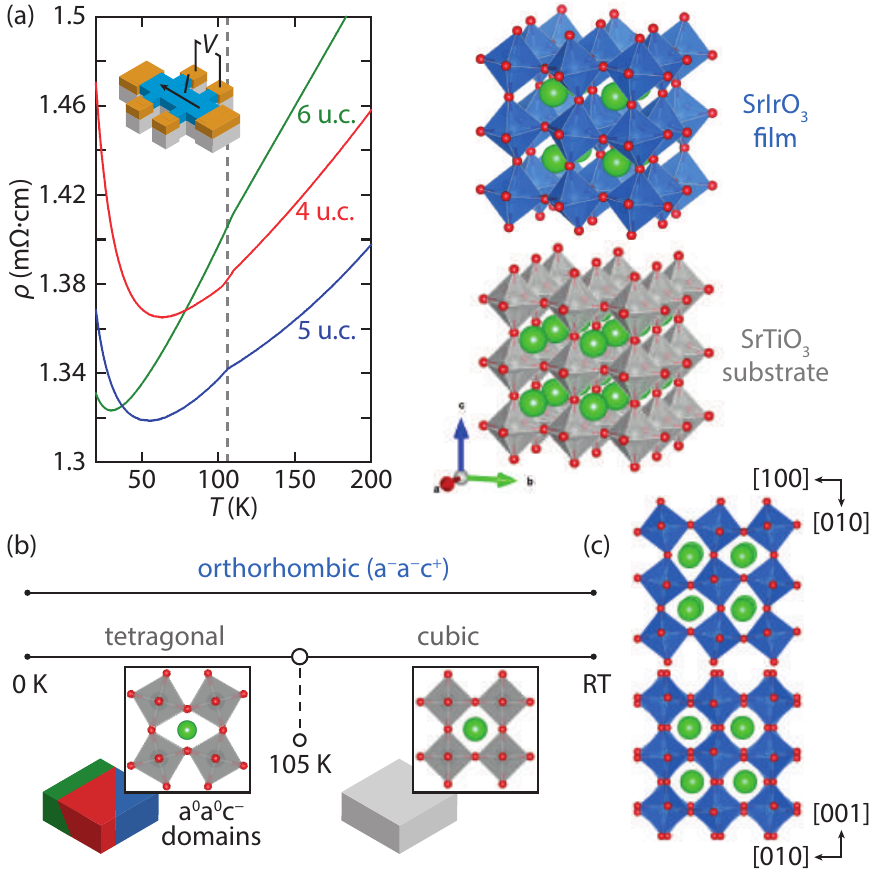}
\caption{\textbf{Simultaneous structural and electronic transition.} (a) $\rho (T)$ curves of \ch{SrIrO3} films of different thicknesses, \textcolor{black}{measured in a HB geometry oriented along the (100) lattice axis.} (b) Bulk phase diagram of \ch{SrIrO3} and \ch{SrTiO3}. \textcolor{black}{Perovskite} \ch{SrIrO3} is orthorhombic at all temperatures while \ch{SrTiO3} undergoes a transition from a cubic to a tetragonal phase below $105\;\textrm{K}$. (c) Octahedral rotations and cation displacements of orthorhombic \ch{SrIrO3} viewed along the pseudocubic [001] (top) and [100] (bottom) directions.}
\label{fig:phase}
\end{figure}

\noindent The resistivity ($\rho$) versus temperature ($T$) characteristics of three \ch{SrIrO3}/\ch{SrTiO3} heterostructures measured in a \textcolor{black}{Hall bar} (HB) geometry are shown in Fig.~\ref{fig:phase}a. The film thicknesses were chosen to be just above the critical point for the metal-insulator transition~\cite{groenendijk2017spin}, such that the properties of the films are most sensitive to interface effects while maintaining a semimetallic ground state. At $T=105\;\textrm{K}$, $\rho$ displays a sudden change of slope. Note that the change in resistivity of the \ch{SrIrO3} film occurs simultaneously with the structural phase-transition in the \ch{SrTiO3} substrate, indicating a strong octahedral connectivity across the interface that couples the lattice degrees of freedom of the \ch{SrTiO3} substrate to the electronic properties of the \ch{SrIrO3} film. The bulk phase diagrams and lattice structures of \ch{SrIrO3} and \ch{SrTiO3} are shown in Fig.~\ref{fig:phase}b and 1c~\cite{momma2011vesta}. \textcolor{black}{Perovskite} \ch{SrIrO3} has an orthorhombic structure (space group \textit{Pbnm}) from $300\;\textrm{K}$ down to low temperature, \textcolor{black}{with rotation angles of typically $10 ^\circ$ or larger about the pseudocubic lattice axes~\cite{zhao2008high, kronbo2016high}.} \ch{SrTiO3} is cubic (\textit{Pm$\bar{3}$m}) but transforms into a tetragonal phase (\textit{I$4$mcm}) below $105\;\textrm{K}$, where it forms three possible domains. Its transition temperature, as well as the magnitude of the distortion can be controlled by e.g. \ch{Ca}- or \ch{Ba}-doping~\cite{bianchi1994raman, guzhva1997spontaneous, menoret2002structural}.

\begin{figure}
\includegraphics[width=\linewidth]{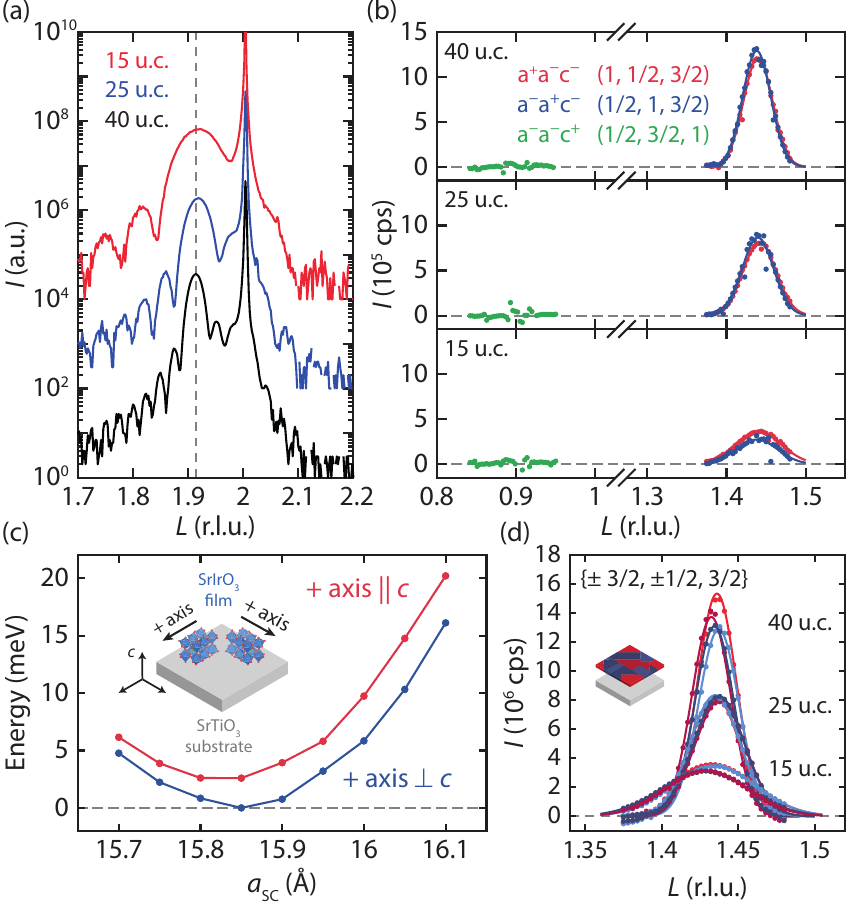}
\caption{\textbf{Binary domain structure.} (a) \textcolor{black}{XRD $L$-scans} of \ch{SrIrO3} films of different thicknesses, measured in the vicinity of the (002) reflection of the \ch{SrTiO3} substrate. (b) Half-order peaks arising from in-phase octahedral rotations. (c) DFT calculated energy difference per formula unit for the in-phase axis (red) parallel and (blue) perpendicular to the $c$-axis (growth axis) as a 
function of lattice constant for supercells consisting of four formula units of \ch{SrTiO3} and \ch{SrIrO3}. (d) Half-order peaks from different rotational domains.}
\label{Fig:XRD1}
\end{figure}

\noindent Octahedral rotations double the perovskite unit cell, a phenomenon that gives rise to half-order Bragg peaks in X-ray diffraction measurements. The presence of specific half-order peaks is governed by symmetry~\cite{glazer1975simple}, and the measurement of a set of half-order peaks can be used to fully determine the rotational pattern of the film~\cite{brahlek2017structural}. \ch{SrTiO3} is characterized by $a^0a^0c^-$ i.e., an out-of-phase rotation about the $c$-axis, which is slightly elongated~\cite{evarestov2011phonon}. Bulk \ch{SrIrO3} is denoted by $a^-a^-c^+$ , having out-of-phase rotations of the same amplitude about two axes and in-phase rotations of different amplitude about the third axis~\cite{liu2016strain}. To study the octahedral rotations in the \ch{SrIrO3}/\ch{SrTiO3} heterostructures, we performed low temperature ($\SI{4}{K}$) synchrotron X-ray diffraction measurements.The films have thicknesses of 40, 25, and 15 u.c.~and are capped by an amorphous \ch{SrTiO3} layer, preventing an additional diffraction signal from the capping layer while shielding the film from exposure to ambient conditions. Measurements of the (002) diffraction peak of these films are shown in Fig.~\ref{Fig:XRD1}a, \textcolor{black}{which demonstrate that} the films are compressively strained. We first consider $(h, k, l)$ Bragg conditions where one of the three reciprocal lattice positions is an integer and the other two are unequal half-order positions $(\nicefrac{1}{2},1,\nicefrac{3}{2})$. This peak is present if the integer reciprocal lattice vector is parallel to the real-space direction of the in-phase axis~\cite{choquette2016octahedral}. As shown in Fig.~\ref{Fig:XRD1}b, a peak is present when the integer reciprocal lattice vector is along $h$ and $k$, but not along $l$. From this we infer that the in-phase rotation (+) axis lies in the plane of the film, and it exhibits a mixed population of $a^+a^-c^-$ and $a^-a^+c^-$ domains, consistent with previous reports~\cite{nie2015interplay,horak2017structure}. In the \ch{ABO3} \textit{Pbnm} structure, the $B$-$B$ distance along the in-phase axis is slightly shorter compared to the out-of-phase axis. Therefore, to minimize the lattice mismatch with the compressive substrate, the in-phase axis should lie in-plane. The $a^-$ axis, which experiences the largest strain, should then be oriented along the $c^-$ axis of \ch{SrTiO3} tetragonal domains, such that $a^-a^+c^-$ ($a^+a^-c^-$) domains in the film couple to $c^-a^0a^0$ ($a^0c^-a^0$) domains in the substrate. This is supported by \textit{ab-initio} calculations (Fig. \ref{Fig:XRD1}c), which show (1) that forming $a^-a^-c^+$ domains is energetically unfavourable due to a larger in-plane lattice parameter when the in-phase axis is oriented out-of-plane ($a_{\text{pc}}=\SI{3.9430}{\angstrom}$) as compared to in-plane ($a_{\text{pc}}=\SI{3.9411}{\angstrom}$)~\cite{zhao2008high} and (2) that the energy is minimized for the aforementioned domain configuration \textcolor{black}{(see Sec. VI of the Supporting Information for further details)}. Different rotational domains arise depending on whether the octahedron closest to the origin rotates clockwise or counterclockwise about each axis. This is probed by the $\{\nicefrac{1}{2},\nicefrac{1}{2},\nicefrac{3}{2}\}$ series of half-order peaks, which provide the $a$ (or $b$) direction along which the displacement of \ch{Sr} ions occurs. Peaks are present for all reflection conditions (Fig.~\ref{Fig:XRD1}d), indicating that the \ch{SrIrO3} film consists of two orthorhombic domains with $a$ aligned along $[100]$ and $[010]$. 

\begin{figure*}
\includegraphics[width=\textwidth]{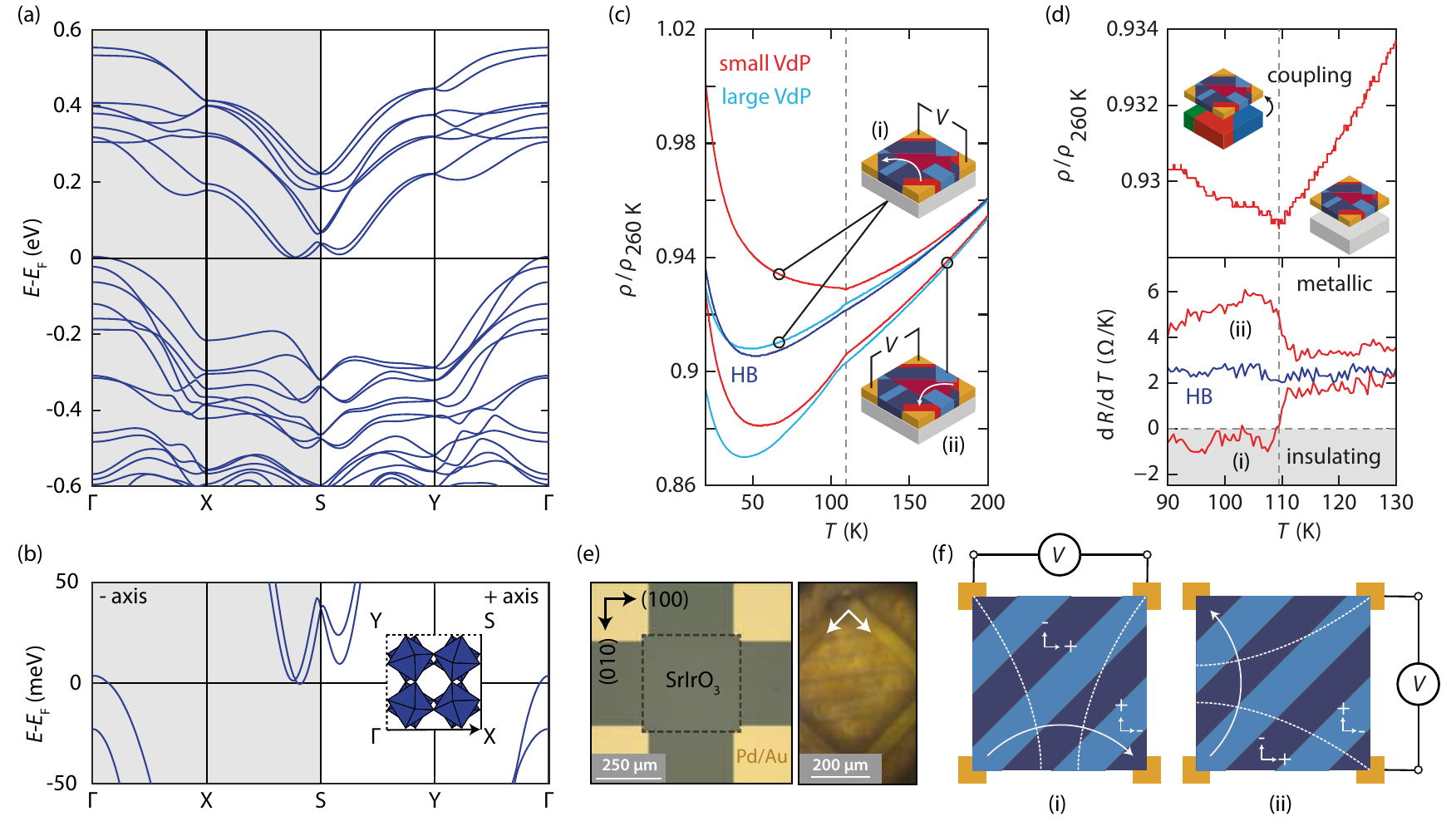}
\caption{\textbf{Anisotropic electronic transport.} (a) DFT-calculated band structure with the out-of-phase ($-$) axis along $\Gamma$--X and the in-phase $(+)$ axis along $\Gamma$--Y. (b) Enlarged view around the Fermi energy. The inset shows the Brillouin zone of the primitive orthorhombic unit cell. (c) $\rho (T)$ curves of a 5 u.c. film comparing (light blue) a large ($\SI{750}{\micro\meter}$) and (red) small ($\SI{375}{\micro\meter}$) VdP geometry, measured in two mutually orthogonal configurations of current and voltage probes. The dark blue curve represents the $\rho (T)$ curve recorded in a \textcolor{black}{$\SI{150}{\micro\meter}$ wide} Hall bar \textcolor{black}{(aspect ratio 3:1)}. (d) Enlarged view of $\rho (T)$ around the cubic-to-tetragonal transition of \ch{SrTiO3} at \SI{105}{K} (top) and the corresponding $\dv*{\rho}{T}$ curves (bottom). (e) Optical microscope images of (left) the $\SI{375}{\micro\meter}$ VdP device and (right) $c^-a^0a^0$ and $a^0c^-a^0$ tetragonal domains in \ch{SrTiO3} \textcolor{black}{in a $\SI{375}{\micro\meter}$ square area}. (f) Illustration of current traversing a binary domain population in the probing region of the device.}
\label{fig:anisotropy}
\end{figure*}

\noindent Having established a coupling between the binary domain structure in the \ch{SrIrO3} film and the tetragonal domains in the \ch{SrTiO3} substrate, we turn to the question of how this interfacial domain coupling affects the electronic properties and the connection with the observed anomaly in the $\rho$-$T$ curve. \textcolor{black}{While in the \textit{Pbnm} structure, the \ch{B - B} distance along the in-phase axis is shorter compared to the out-of-phase axis, the \ch{B - O - B} bond angles are slightly more tilted~\cite{zhou2008intrinsic}.} Accordingly, one would expect a reduction of bandwidth along the in-phase axis due to a reduced orbital overlap~\cite{gruenewald2014compressive}, with anisotropic transport properties as a consequence. Fig.~\ref{fig:anisotropy}a and~\ref{fig:anisotropy}b show the DFT-calculated electronic structure, assuming a correlation strength $U=\SI{1.47}{eV}$, similar to previous work~\cite{groenendijk2017spin}. The out-of-phase (-) axis is oriented along $\Gamma$--X and the in-phase (+) axis along $\Gamma$--Y, with $\Gamma$ the center of the primitive orthorhombic Brillouin zone. Electron wavepackets along $\Gamma$--X have a group velocity oriented purely along the out-of-phase axis and along X--S include a component along the in-phase axis, which is smaller than or equal to the component along the out-of-phase axis. Accordingly, $\Gamma$--X--S (gray region) comprises carrier transport oriented either fully or predominantly along the out-of-phase axis (and analogously for S--$\Gamma$--Y and the in-phase axis). Two electron-like pockets are present along X--S and S--Y. However, only the former intersects the Fermi level and the latter remains unoccupied. As a consequence, electronic bands along the in-phase axis are depleted at the Fermi level and the system is anticipated to favour insulating behaviour along the in-phase axis, but remain metallic along the out-of-phase axis. This is a remarkable scenario, where the electronic structure is finely tuned between a metallic and insulating phase by a reduction of bandwidth along the in-phase axis. In \ch{AIrO3} iridates, time-reversal symmetry protects the nodal line and thus safeguards metallic behaviour. A metal--insulator transition therefore necessarily coincides with the onset of $G$-type antiferromagnetic order~\cite{kim2015surface, matsuno2015engineering, groenendijk2017spin}. Our DFT calculations confirm that AFM is required to realize any type of insulating behaviour, even if it is anisotropic in nature. Experimentally, we indeed observe strongly anisotropic electronic properties. Fig.~\ref{fig:anisotropy}c shows $\rho (T)$ measured in a HB geometry and in two patterned \textcolor{black}{van der Pauw (VdP)} squares with sizes of 375 and $\SI{750}{\micro\meter}$ for two electrical configurations. We directly observe that the anomaly in $\rho$ is much more pronounced in the VdP geometry than in the Hall bar and that a strong anisotropy develops below $\SI{105}{K}$. As shown in Fig.~\ref{fig:anisotropy}d, the transition can be remarkably sharp and manifest as a metal-insulator transition. The derivative $\textrm{d}\rho/\textrm{d}T$ is shown in the bottom panel, which shows opposite behaviour in the two electrical configurations i.e., a positive (metallic) or negative (insulating) slope depending on the orientation. Microscopically, this can be viewed as current traversing an unequal domain population in the probing region of the VdP device (see Figs.~\ref{fig:anisotropy}e and \ref{fig:anisotropy}f). Domains in \ch{SrTiO3} can be sized up to $\SI{100}{\micro\meter}$ (see also Sec.~IV of the Supporting Information), which suggests, in accordance with our observations, that the anisotropic character should be most pronounced in small devices and reduced in larger devices due to statistical averaging over complex domain patterns~\cite{honig2013local, kalisky2013locally,frenkel2016anisotropic}. The $\rho (T)$ anomaly at 105 K can then be ascribed to a sudden reconfiguration of the current paths as the \ch{SrIrO3} domains adapt to the onset of the tetragonal multi-domain state of the \ch{SrTiO3} substrate. \textcolor{black}{We remark that at the boundaries between adjacent structural domains, the crystal unit cells are typically distorted~\cite{honig2013local}. Considering the strong structure-property relationship in iridates, it is likely that the domain walls have different electronic properties compared to the undistorted areas. However, due to the $\sim \SI{45}{\degree}$ angle with respect to the crystal lattice axes, any enhanced or suppressed conductivity would be projected equally onto the (100) and (010) directions. Hence, the devices shown in Fig.~\ref{fig:anisotropy} are only sensitive to the domains and not to the domain walls. Probing transport in nanoscale devices oriented at $\SI{45}{\degree}$ could elucidate their electronic properties.} 


\begin{figure}
\includegraphics[width=\linewidth]{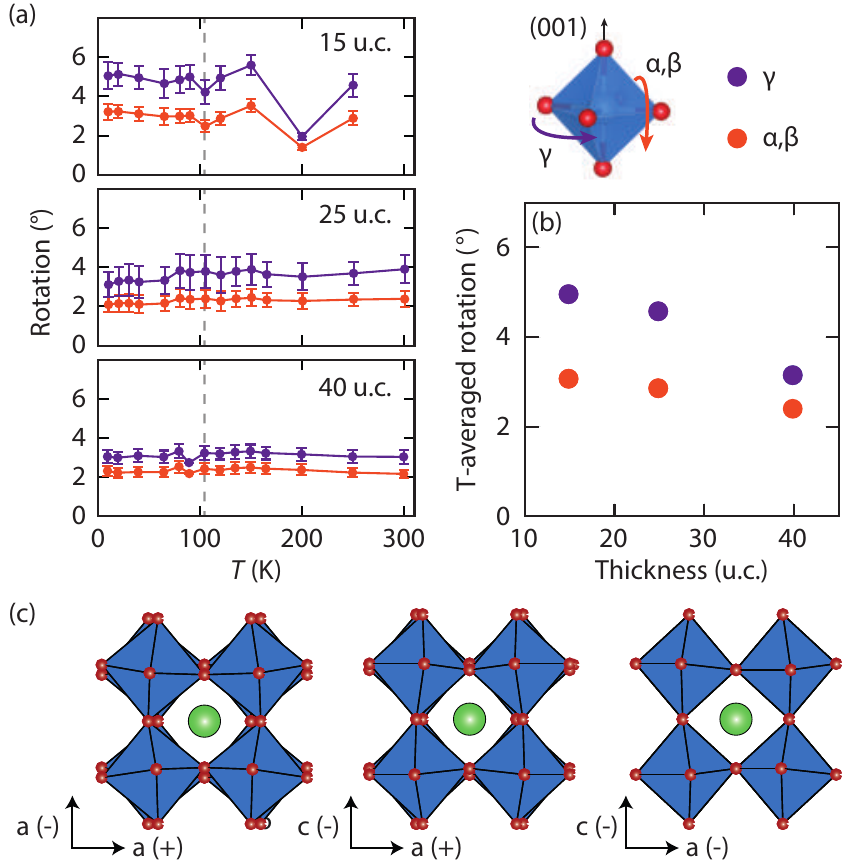}
\caption{\textbf{Temperature dependence of octahedral rotations.} (a) Rotation angles of the 15, 25, and 40 u.c.~films as a function of temperature. (b) Temperature-averaged rotation angles as a function of film thickness. (c) Visualization of the octahedral rotation pattern as seen (from left to right) along the $c^-$, $a^-$ and $a^+$ axes, respectively. }
\label{Fig:XRD2}
\end{figure}

\noindent To further explore the effect of the \ch{SrTiO3} tetragonal distortion on the octahedral rotations in \ch{SrIrO3}, we performed temperature-dependent diffraction measurements across the transition temperature (see Fig.~\ref{Fig:XRD2}). By fitting the half-order Bragg peaks with a Gaussian function and comparing the areas under the curves, we quantify the octahedral rotation angles and cation displacements as a function of temperature~\cite{may2010quantifying}. The oxygen positions are obtained by comparing the intensities of the peaks with the calculated structure factor of the oxygen octahedra. \textcolor{black}{Standard nonlinear regression is used to determine the optimal values of $\alpha$ and $\gamma$, defined in Fig.~\ref{Fig:XRD2}. The determined in- and out-of-plane rotation angles $\alpha$ and $\gamma$, respectively, are plotted versus temperature for \ch{SrIrO3} films of different thicknesses in Fig.~\ref{Fig:XRD2}a.} The angles are found to be nearly constant over the entire temperature range and weakly dependent on the film thickness (Fig.~\ref{Fig:XRD2}b). Fig.~\ref{Fig:XRD2}c visualizes the low temperature lattice structure. The rotational angles are substantially reduced with respect to bulk \ch{SrIrO3}. Considering that \ch{SrTiO3} has been reported to strongly suppress octahedral rotations in other oxide heterostructures~\cite{liao2016controlled}, we attribute this to the interaction with the \ch{SrTiO3} substrate~\textcolor{black}{\cite{guo2020engineering}}. \textcolor{black}{We also find an enhancement of orthorhombicity for the thinner films, possibly pointing to larger rotational distortions in the unit cells closest to the \ch{SrTiO3}/ \ch{SrIrO3} interface (see also Sec. VI of the Supporting Information). Interestingly, we do not observe a clear deviation of the \ch{SrIrO3} rotation angles across $\SI{105}{K}$, further pointing to the reconfiguration of the multi-domain state as the underlying cause of the observed resistivity anomaly at 105 K.}


\noindent In summary, we established an interfacial coupling in ultrathin \ch{SrTiO3}/\ch{SrIrO3} heterostructures and demonstrated the emergence of a binary orthorhombic domain pattern in \ch{SrIrO3} that couples directly to the tetragonal domains in the \ch{SrTiO3} substrate. For each domain, the electronic bandwidth along the in-phase rotational axis is suppressed, resulting in strongly anisotropic transport properties that manifest as a metal--insulator transition. This coupling mechanism is not limited to iridates, but can be extended to control physical properties such as magnetism, multiferroicity and superconductivity in a wide variety of orthorhombic materials, e.g.  ferrites, manganites and nickelates~\cite{liu2013strain, bhattacharjee2009engineering, bousquet2016non, gibert2015interfacial,caviglia2012ultrafast, fowlie2019thickness, li2019superconductivity}.\\


\noindent \textbf{Experimental Methods.} \ch{SrIrO3} thin films were synthesized by pulsed-laser \textcolor{black}{deposition} on (001) \ch{TiO2}-terminated \ch{SrTiO3} substrates. \textcolor{black}{The growth conditions are described in detail in previous work, including the requirement of a protective capping layer to prevent degradation of the films resulting from exposure to ambient conditions~\cite{groenendijk2016epitaxial}. Samples measured in transport were capped by a 10 u.c. crystalline layer of \ch{SrTiO3}, whereas samples measured in XRD were capped by amorphous \ch{SrTiO3}, to prevent an additional contribution in diffraction.} Hall bar (HB) and van der Pauw (VdP) geometries were patterned by e-beam lithography. The \ch{SrIrO3} layer was contacted by Ar etching and \textit{in-situ} deposition of Pd and Au, \textcolor{black}{resulting in low-resistance Ohmic contacts (see also Section V.A of the Supporting Information)}. Low temperature transport measurements were performed in an Oxford flow cryostat, \textcolor{black}{by sourcing a low frequency ($\sim\SI{17}{Hz}$) $\SI{10}{\micro A}$ current and measuring the resulting voltage with a lock-in amplifier}. Details regarding the synchrotron X-ray diffraction measurements, half-order peak analysis, \textcolor{black}{polarized-light microscopy measurements} and \textcolor{black}{\textit{ab-initio}} calculations are described in the Supporting Information.\\


\noindent \textcolor{black}{\textbf{Supporting Information.} The Supporting Information contains (I) details regarding the determination of the octahedral rotation angles, (II) the observation of rotational distortions in \ch{SrTiO3} above the condensation point, (III) additional diffraction measurements on ultrathin \ch{SrIrO3} films, (IV) imaging of tetragonal domains in \ch{SrTiO3}, (V) details regarding the device fabrication, as well as additional transport measurements on a 30 u.c. film and (VI) additional \textit{ab-initio} calculations. The crystallographic data from the DFT structural relaxation is included as the file labeled 'POSCAR'.} \\


\noindent \textbf{Acknowledgements.} The authors acknowledge D.~J.~Groenendijk, M.~Lee, P.~Willmott and D.~Porter for experimental support and discussions. The authors thank S.~Picozzi for discussions and collaboration and P.~G.~Steeneken and H.~S.~J.~van der Zant for use of equipment. The authors are grateful to J.~R.~Hortensius, J.~de~Bruijckere and H.~Thierschmann for input on the manuscript. \\
\noindent This work was supported by the European Research Council under the European Unions Horizon 2020 programme/ERC Grant agreements No. [677458] and No. [731473] (Quantox of QuantERA ERA-NET Cofund in Quantum Technologies) and by the Netherlands Organisation for Scientific Research (NWO/OCW) as part of the Frontiers of Nanoscience program (NanoFront) and VIDI program. This work was supported by the Swiss National Science Foundation - division II and has received funding from the European Research Council under the European Union Seventh Framework Programme (No. FP7/2007-2013)/ERC Grant Agreement No. 319286 (Q-MAC). C.A. acknowledges support from the Foundation for Polish Science through the IRA Programme co-financed by the EU within SG OP. This research was carried out with the support of the Interdisciplinary Centre for Mathematical and Computational Modelling (ICM) University of Warsaw under Grant No. G73-23 and G75-10. This work was carried out with the support of the Diamond Light Source Beamline I16 (Didcot, UK) and the MS - X04SA Materials Science beamline at the Paul Scherrer Institute (Villigen, Switzerland).

\balance{\bibliography{References.bib}}



\section*{Supplementary material}

# Supplementary Information for:

# Coupling lattice instabilities across the interface in ultrathin oxide heterostructures

T.C. van Thiel,[1] J. Fowlie,[2] C. Autieri,[3,4] N. Manca,[1]
M. Šiškins,[1] D. Afanasiev,[1] S. Gariglio,[2] and A. D. Caviglia[1]

[1] *Kavli Institute of Nanoscience, Delft University of Technology,
Lorentzweg 1, 2628 CJ Delft, Netherlands*
[2] *Department of Quantum Matter Physics, University of Geneva,
24 Quai Ernest-Ansermet, 1211 Genève 4, Switzerland*
[3] *International Research Centre MagTop, Institute of Physics,
Polish Academy of Sciences, Aleja Lotników 32/46, PL-02668 Warsaw, Poland*
[4] *Consiglio Nazionale delle Ricerche, Istituto Superconduttori,
Materiali Innovativi e Dispositivi (CNR-SPIN),
c/o Università G. D'Annunzio, I-66100 Chieti, Italy*

**CONTENTS**

## I. DETERMINATION OF THE OCTAHEDRAL ROTATIONS

The X-ray diffraction measurements were carried out at the I16 beamline at Diamond Light Source in the form of a series of off-specular crystal truncation rods (CTRs), centered on half-integer Bragg diffraction positions. The CTRs were recorded on a Pilatus 100k photon-counting pixel detector, at a fixed incidence angle of 4° and a photon energy of 8 keV. For each sample, the same series of CTRs was recorded at temperatures ranging from 10 to 300 K. We find that that all films, at all temperatures, adhere to the *Pbnm* symmetry with the long (in-phase rotation) orthorhombic axis always perpendicular to the growth axis i.e., in both $a$ and $b$ directions. In other words, the system can be described in Glazer notation as a combination of $a^+a^-c^-$ and $a^-a^+c^-$ [1]. The population fractions of these domains can be estimated by comparing the intensities of e.g. $(1/2, 1, 3/2)$ and $(1, 1/2, 3/2)$, which were found to be approximately equal in the area of the beam spot. Rotations about the $a$ and $b$ axes, whether they be out-of-phase or in-phase, are assumed to be equal in magnitude given the approximate square in-plane symmetry of the heterostructure. Finally, four geometric domains (oriented along $[100], [\bar{1}00], [010]$ and $[0\bar{1}0]$) are found to exist in approximately equal proportion from observing the presence of equally intense Bragg peaks that belong to the same family, such as $(3/2, 1/2, 3/2)$, $(-3/2, 1/2, 3/2)$, $(3/2, -1/2, 3/2)$ and $(-3/2, -1/2, 3/2)$, where going from one to the next corresponds simply to a rotation of the sample about its own normal by 90°. Quantification of the tilt and rotation angles is further garnered from the intensities of the allowed half-integer Bragg peaks in the manner first introduced by May and coworkers and employed successfully in many instances since [2, 3]. The experimental integrated intensity, $I_{\text{exp}}$, was extracted from each CTR and compared to a simulated diffraction intensity, $I_{\text{sim}}$, for the same $(h, k, l)$ and for a given set of Glazer angles, $\alpha$, $\beta$ and $\gamma$. Then, upon varying the input angles, the best fit simulated structure is found by the minimizing the residual sum of squares (RSS), defined as

$$\text{RSS} = \sum_n \left| I_{\text{sim}}(h,k,l)_n - I_{\text{exp}}(h,k,l) \right|^2 \tag{1}$$

with

$$I_{\text{sim}}(h,k,l) = \sum_{v=1}^{2} \sum_{p=1}^{4} \left[ f_{O^{2-}} \sum_{n=1}^{24} \exp\left(2\pi i \left(hx_{npv} + ky_{npv} + lz_{npv}\right)\right) \right] \tag{2}$$

The simulated diffraction intensity is calculated from the square of the structure factors for the 24 $O^{2-}$ ions that constitute the perovskite pseudocubic unit cell of $SrIrO_3$ (SIO) doubled in all three directions. The two equally populated orthorhombic domains as well as the four equally populated geometric domains must also be summed over, represented by the indices v and p respectively. The atomic positions, $(x_{nvp}, y_{nvp}, z_{nvp})$ for each of the n oxygen atoms used to generate the structure factors are obtained after careful application of three dimensional rotation matrices to an undistorted octahedron, as described in ref. [4]. This method permits determination of Sr-cation displacements, however the best fitting results were obtained by locking them to a cubic sublattice. Fig. S1 shows an example of a fitting result for a 25 u.c. film, comparing temperatures 250 K and 10 K. As discussed in the main text, the rotation angles are nearly independent of temperature and significantly reduced with respect to bulk SIO due to the strain exerted by the $SrTiO_3$ (STO) substrate. In Fig. S2, we show a subset of the fitted reflections for three samples of varying thicknesses, comparing low (10 K) and high temperature (250 K). The thinnest films show the largest rotation angles, which may be attributed to the symmetry mismatch between the SIO (SIO) film and STO substrate, causing larger orthorhombic distortions in the near-interface region.

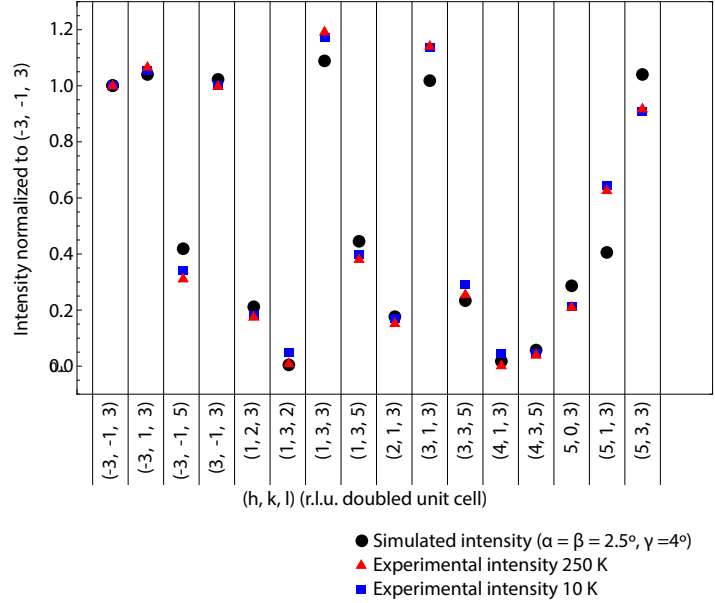

FIG. S1: Experimental and simulated half-order peak intensities of a 25 u.c. SIO film for temperatures 10 K and 250 K. The black dots indicate simulated half-order peak intensities for $\alpha = \beta = 2.5°$ and $\gamma = 4°$.

## II. ROTATIONAL DISTORTIONS IN STO ABOVE THE CONDENSATION POINT

Fig. S3 shows XRD $L$-scans in the vicinity of the STO $(3/2, 1/2, 5/2)$ reflection for a 30 u.c. SIO film. This reflection probes out-of-phase rotation about the axis normal to the film plane ($a^0 a^0 c^-$ in the case of STO). We first focus on the curve at 20 K, which is well below the cubic-to-tetragonal transition temperature of STO. The intense peak at $L = 2.5$ originates from the substrate, whereas the broader peak situated slightly below $L = 2.4$ originates from the SIO film. As the temperature is increased, the substrate contribution is seen to diminish, but does not vanish entirely. For temperatures 120 K and above, the sharp feature is no longer present but a broader peak around $L = 2.5$ remains discernible up to 300 K. This indicates the presence of tetragonality in STO well above 105 K. As discussed in the main text, the strong octahedral connectivity across the STO/SIO interface causes the STO substrate to force a rotational pattern on the SIO film. Such a coupling can work both ways, meaning that the SIO film may also promote rotations in the top layers of the substrate (this is further discussed in Section VI C). Alternatively, this could be the result of the STO surface layers behaving differently from the bulk [5, 6]. We remark that the degree of tetragonality in STO above 105 K is small and that macroscopic condensation of the entire substrate into a tetragonal state, coinciding with a redistribution of the domain configuration and current paths in the SIO film, only occurs at 105 K. Nevertheless, it is possible that it contributes to the pinning of domains, meaning there is a larger probability of encountering a similar (or identical) domain structure in STO (and possibly SIO) across different heating/cooling cycles. This could explain the degree of anisotropy that is already present above 105 K in Fig. 3c of the main text.

## III. DIFFRACTION MEASUREMENTS ON ULTRATHIN FILMS

In Fig. S4, we show diffraction data on a 10 u.c. and 4 u.c. film, comparing half-order peak intensities at 300 K and 20 K. We find that for the 10 u.c. film, some structural differences between 300 K and 20 K are present, but subtle. For the 4 u.c. film, not all reflections had a sufficient photon

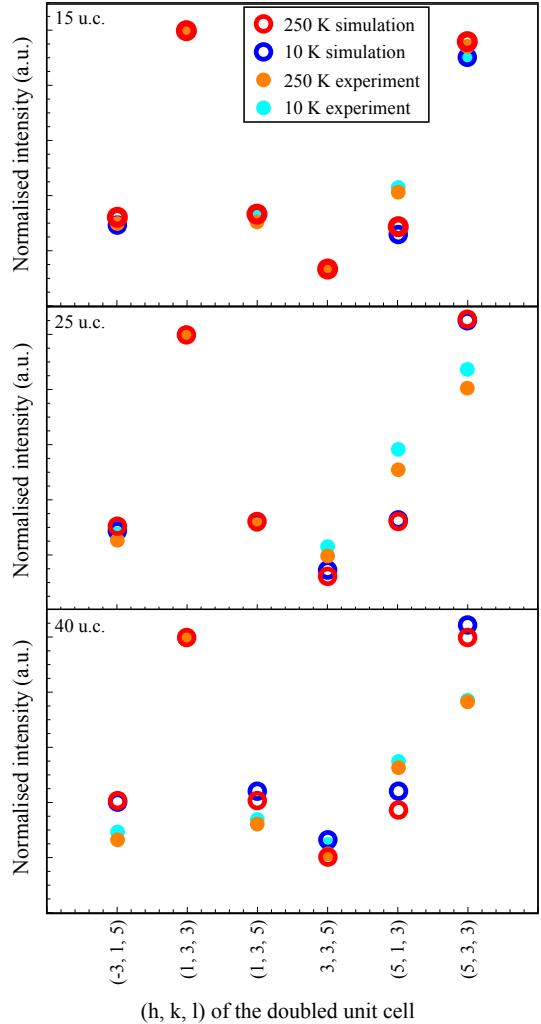

FIG. S2: Simulated half-order peak intensities for a selected subset of reflections recorded at 10 K and 250 K, comparing samples of three different thicknesses. The open circles represent the best fit simulation intensities at the relevant temperature. The simulated rotation angle about the c-axis $\gamma$ is indicated in blue and simulated angles about the in-plane lattices axes ($\alpha = \beta$) are indicated in red. The angles obtained from the fitting procedure are; 15 u.c. (10 K) $\alpha = \beta = 3.2°$, $\gamma = 5.0°$ and (250 K) $\alpha = \beta = 2.9°$, $\gamma = 4.5°$; 25 u.c. (10 K) $\alpha = \beta = 2.0°$, $\gamma = 3.1°$ and (250 K) $\alpha = \beta = 2.4°$, $\gamma = 3.8°$; 40 u.c. (10 K) $\alpha = \beta = 2.3°$, $\gamma = 3.0°$ and (250 K) $\alpha = \beta = 2.2°$, $\gamma = 3.0°$.

count to perform a complete analysis. Nevertheless, the comparison between high and low temperature in panel (b) reveals a significant change of the $(1.5, 0.5, 1)$ reflection at low temperature, suggesting a more pronounced change in rotation angles of ultrathin SIO across the STO phase transition. As discussed in Section II, the peak observed at 300 K at $L = 1.5$ corresponds to the STO substrate and is indicative of tetragonal distortions in STO above the transition point. These observations are consistent with a gradient of rotation angles along the growth axis. Additional *ab-initio* calculations support this picture (discussed in Section IV.B).

## IV. IMAGING OF STO TETRAGONAL DOMAINS

Below 105 K the structure of STO is split into three types of tetragonal domains, each of them corresponding to a rotation and elongation of the unit cell along one of three equivalent directions.

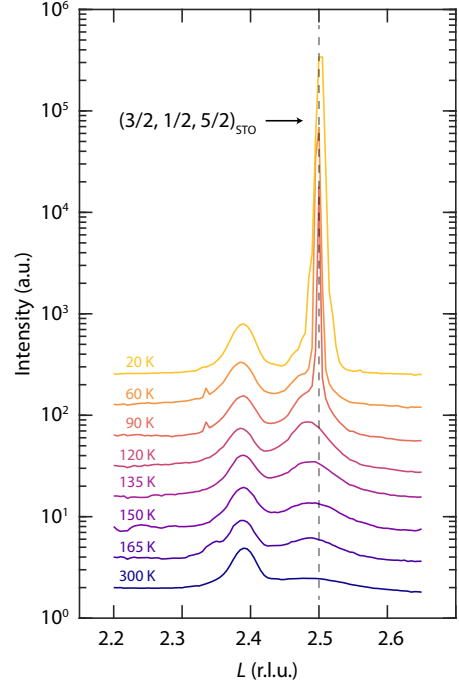

FIG. S3: $L$-scan in the vicinity of the $(3/2, 1/2, 5/2)$ reflection of STO.

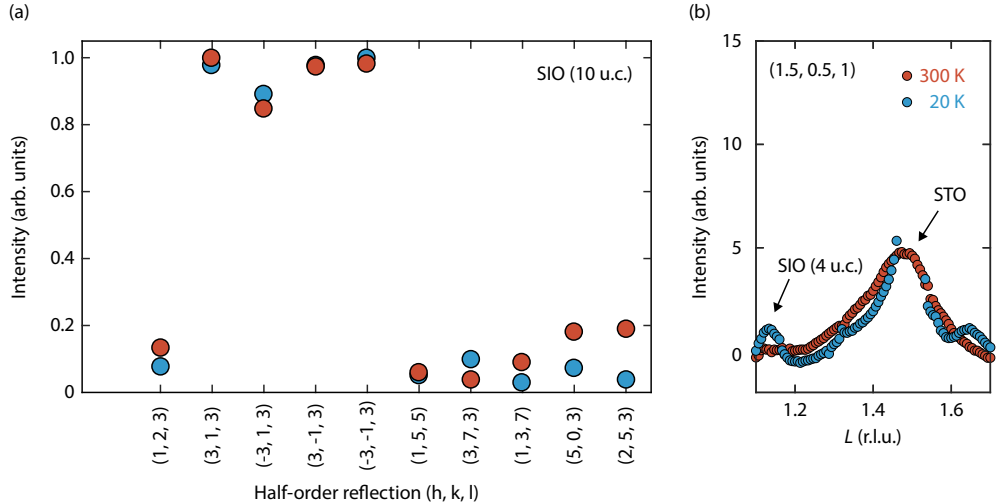

FIG. S4: (a) Half-order peak intensities for a 10 u.c. SIO film on STO measured at (red) 300 K and (blue) 20 K. The intensity values have been normalized to the (3,1,3) peak intensity. Panel (b) shows an $L$-scan in the vicinity of the $(1.5, 0.5, 1)$ reflection for a 4 u.c. SIO film, comparing high (300 K, red) and low temperature (20 K, blue).

In the main text, we demonstrated that to minimize the lattice mismatch, $c^-a^0a^0$ and $a^0c^-a^0$ domains in the STO substrate are coupled to $a^-a^+c^-$ and $a^+a^-c^-$ domains in the overlying SIO film, respectively. Since tetragonal domains in STO are birefringent [7], this provides an opportunity to optically study the multi-domain state. Here we use polarized-light microscopy to visualize the domain structure of an STO substrate. The light from an LED source was collimated, polarized with a Glan-Taylor prism and focused on the sample surface by an optical objective. The light reflected from the sample was collected by the same objective and directed to the sensor area of a digital camera. The polarization contrast was acquired by placing a second polarizer (analyzer)

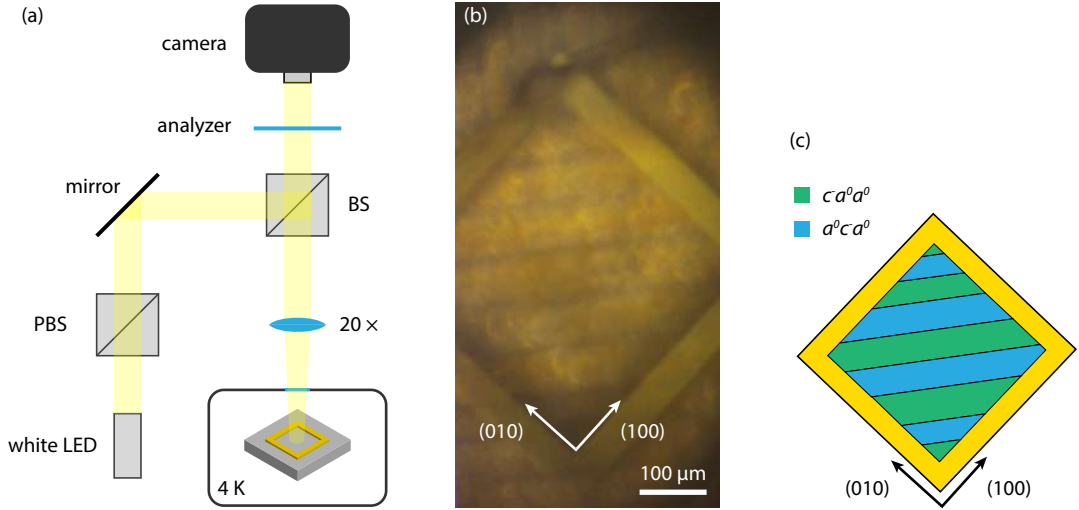

FIG. S5: (a) Schematic illustration of the polarized light microscopy setup. (b) image recorded with crossed polarizers, showing STO tetragonal domains and (c) illustration of (b) showcasing the different domains joined by twin boundaries.

in the optical path of the reflected beam such that the mutual polarizations of the two polarizers were nearly orthogonal (cross-Nikol configuration). To maximize the signal, the measurements were done at the lowest accessible stable temperature of 4 K. As discussed in ref. [8], $a^0c^-a^0$ and $c^-a^0a^0$ domains in STO are joined by twin boundaries oriented at approximately 45° with respect to the (100) or (010) lattice axis. In the vicinity of the domain boundary, the STO unit cells are distorted, resulting in a locally modified birefringence. Accordingly, in the cross-Nikol configuration, the tetragonal domains appear bright and the boundaries between them appear dark.

In Fig. S5b, we show a microscopy image of the polarization contrast obtained in the cross-Nikol configuration. To facilitate comparison with the devices discussed in the main text, a Cr/Au open square geometry of inner width 375 μm is patterned on top. Nine different domains are discernible within the square area. Fig. S5c shows a trace of the domains. Note that this experiment only identifies the domain boundaries and not the domains themselves. The illustration in (c) represents therefore only one of two possible configurations. The relatively large size of the domains with respect to the device dimensions leads to an increased probability of encountering an unequal distribution of domain areas. In this particular example, we find a distribution of four versus five with the majority domain type having of about 4% larger surface area within the square. As the probing area of a van der Pauw device is smaller than the full square area, the imbalance probed in a transport experiment is statistically likely to be larger. In the scenario predicted by the *ab-initio* calculations that one domain type is significantly more conductive than the other, this percentage matches well with the experimentally observed resistivity anisotropy in Fig. 3c of the main text.

## V. TRANSPORT MEASUREMENTS

### A. Device Fabrication

The SIO thin films are grown by pulsed-laser deposition (see ref. [9] for a detailed description of the growth procedure). To prevent degradation of the films resulting from exposure to ambient conditions, the samples were capped by a protective STO layer. The samples measured in

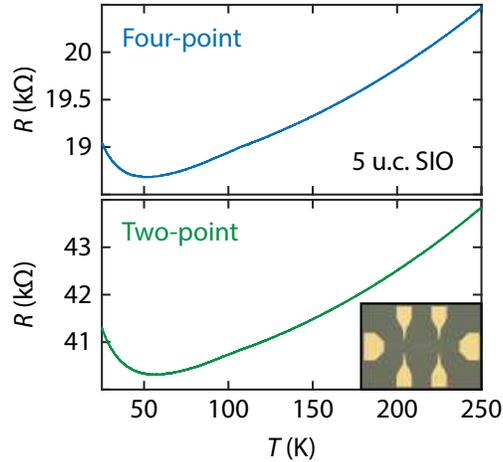

FIG. S6: **Ohmic contacts.** (Top) four-point and (bottom) two-point resistance as a function of temperature for a 5 u.c. SIO film measured in a 150 μm wide Hall bar geometry (see inset). The distance between the two voltage probes is 450 μm, yielding an aspect ratio of 3:1.

XRD were capped by amorphous STO (to prevent an additional contribution in diffraction), while samples measured in transport were capped by 10 u.c. of crystalline STO. Device fabrication is performed with a two-step e-beam lithography recipe, using PMMA as a positive resist (495 kDa A8, dose 800 μC/cm$^2$). In the first step, the metal contact areas are exposed. Embedded contacts are created by Ar-milling, followed by in-situ evaporation of Pd and Au. The second e-beam step defines the device geometries. The conductive SIO film is removed from the non-device areas through Ar-milling, using PMMA as an etch mask. The four-point and two-point resistances of a 5 u.c. SIO film measured in a HB geometry are shown in Fig. S6. Both curves show qualitatively the same behavior, indicating that the contacts are Ohmic throughout the full temperature range. The approximate factor of 2 difference between the resistance values arises primarily from the larger aspect ratio of the two-point geometry compared to the four-point geometry.

## B. Resistivity anomaly in a 30 u.c. SIO film

In Fig. S7, we show transport measurements of a 30 u.c. SIO film (see also ref. [9]). The full resistance curve as a function of temperature is shown in Fig. S7a, whereas Fig. S7b presents an enlarged view around the cubic-tetragonal transition of the STO substrate. While not as pronounced as in the thinner films, an anomaly is observed at the transition, also identified from the derivative with respect to temperature shown in panel Fig. S7c. The underlying reason for the less pronounced features in thicker films is their more strongly metallic character. While an anisotropy is still present, the electron pockets along both the in- and out-of-phase axis intersect the Fermi level, rendering the anisotropic properties less prominent. A more detailed discussion on the electronic properties of SIO thin films is presented in Section VI B.

8

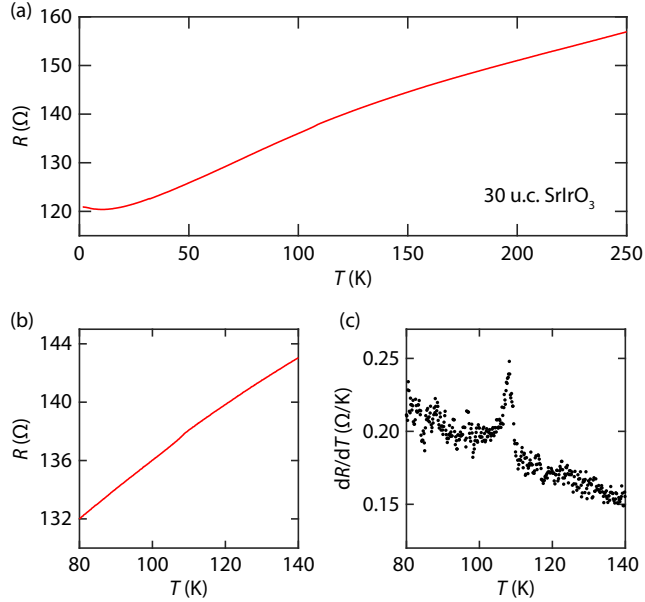

FIG. S7: **Resistivity anomaly.** Transport measurements on a 30 u.c. SIO film, showing the resistance as a function of temperature in (a) a wide temperature range and (b) an enlarged view in the vicinity of 105 K. Panel (c) shows the derivative $\mathrm{d}R/\mathrm{d}T$ corresponding to (b).

## VI. AB-INITIO CALCULATIONS

### A. Methodology

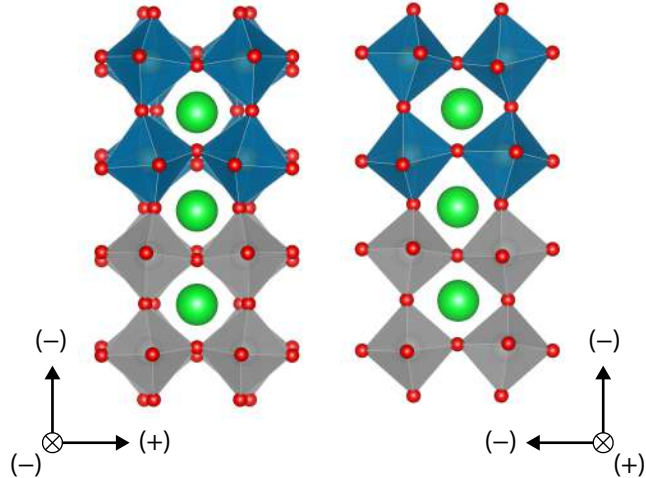

FIG. S8: **Structural relaxation.** STO/SIO lattice structure following from the structural relaxation, viewed along (left) the out-of-phase axis and (right) the in-phase axis.

Ab-initio density functional theory calculations were carried out within the generalized gradient approximation (GGA), by using the plane-wave VASP package and the PBEsol for the exchange-correlation functional with spin-orbit coupling [10, 11]. Computations were performed for a supercell with 4 formula units of STO and 4 formula units of SIO with the in-phase axis (+ axis) parallel to the growth axis, as well as a supercell with 8 formula units of STO and 8 formula units of SIO with the + axis perpendicular to the growth axis. A $8 \times 8 \times 2$ and $6 \times 6 \times 2$ k-point Monkhorst-Pack

grid was used for all calculations on the supercells of 40 and 80 atoms, respectively. The structural relaxation was performed separately for each volume. Hubbard $U$ effects between Ir sites were included within the GGA + U [12], with a value $J_\text{H} = 0.15U$ for the Hund's coupling. Experimentally observed insulating behaviour is reproduced by introducing $G$-type antiferromagnetic order with $U = 1.47\,\text{eV}$. This value is close to the previously reported value for antiferromagnetic SIO [13]. A visualization of the STO/SIO heterostructure resulting from the structural relaxation is shown in Fig. S8. The out-of-phase axes of STO and SIO are aligned, preserving the symmetry continuity across the interface.

### B. Effect of symmetry breaking on the electronic structure of ultrathin SIO

Due to protection of the Dirac nodal ring, the electronic structure of SrIrO$_3$ is always metallic when the lattice and time reversal symmetry are preserved [14, 15]. To realize an insulating ground state, the Dirac nodal ring must be gapped, which occurs when one of the aforementioned symmetries is broken [16, 17]. At the SIO/STO interface, the lattice symmetry is broken, but the system retains metallicity. The key ingredient is a strong on-site Coulomb repulsion $U$. When $U$ dominates over the electronic bandwidth $W$, time-reversal symmetry is broken and an antiferromagnetic insulating state is realized. In the ultrathin limit, out-of-plane hopping is suppressed and $W$ is reduced. As the $U/W$ ratio increases, the system experiences a metal-insulator transition. In previous work, we showed that the critical thickness for metallicity in ultrathin SIO is between 3 and 4 u.c. with an on-site Coulomb repulsion of $1.5\,\text{eV}$ [13]. The heterostructure considered in the present work is 4 u.c. thick and is therefore situated precisely at the boundary between metallic and insulating. As demonstrated in the main text, the SIO in-phase rotational axis is perpendicular to the growth axis in STO/SIO heterostructures, which means that the in-plane lattice symmetry is broken. This is included in the *ab-initio* calculations and is immediately apparent from Fig. S9a, where the electronic structure exhibits a degree of asymmetry in the S point. We observe a lifting of the electron pocket in the S—Y section above the Fermi energy, favouring an insulating state. Fig. S9b shows the electronic structure for the same lattice, but the on-site Coulomb repulsion set to zero. The asymmetry is still present but both pockets intersect the Fermi energy, producing a metallic state along both the in-phase and out-of-phase axis. From this we infer that the reduced bandwidth in the ultrathin limit is by itself not sufficient to reproduce experimentally observed insulating behaviour and a strong on-site Coulomb repulsion is required. On top of this, the in-plane lattice symmetry breaking is the key ingredient to reproduce experimentally produced the anisotropic transport characteristics. This is an intrinsic property of the orthorhombic *Pbnm* symmetry group due to the presence of two out-of-phase rotations and one in-phase rotation. The in-plane symmetry breaking is enhanced in STO/SIO heterostructures due to the suppression of $a^-a^-c^+$ domains as discussed in the main text. In Fig. S10a, we show the electronic structure with the hypothetical scenario of the in in-phase axis being out-of-plane and the two out-of-phase axes in-plane. As expected, this scenario shows a more isotropic behaviour. Fig. S10b shows the electronic structure with zero on-site Coulomb repulsion, which is seen to reduce the electron effective mass and enhance metallicity.

### C. Influence of strain, octahedral connectivity and confinement on rotation angles

Experimentally the lattice structure of SIO is found to differ from the bulk structure when synthesized in ultrathin form on STO. In this section, we theoretically investigate a number of mechanisms that may alter the octahedral rotation angles of ultrathin SIO/STO heterostructures with respect to bulk SIO. The first scenario we consider is an overall reduction of the SIO unit cell

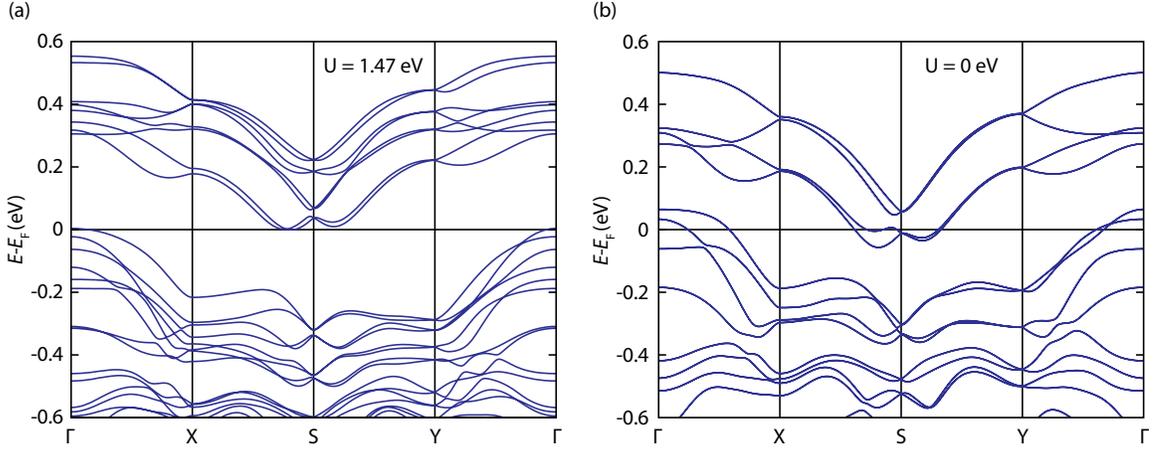

FIG. S9: SIO/STO electronic structure with the in-phase axis in-plane (a) with and (b) without on-site Coulomb repulsion.

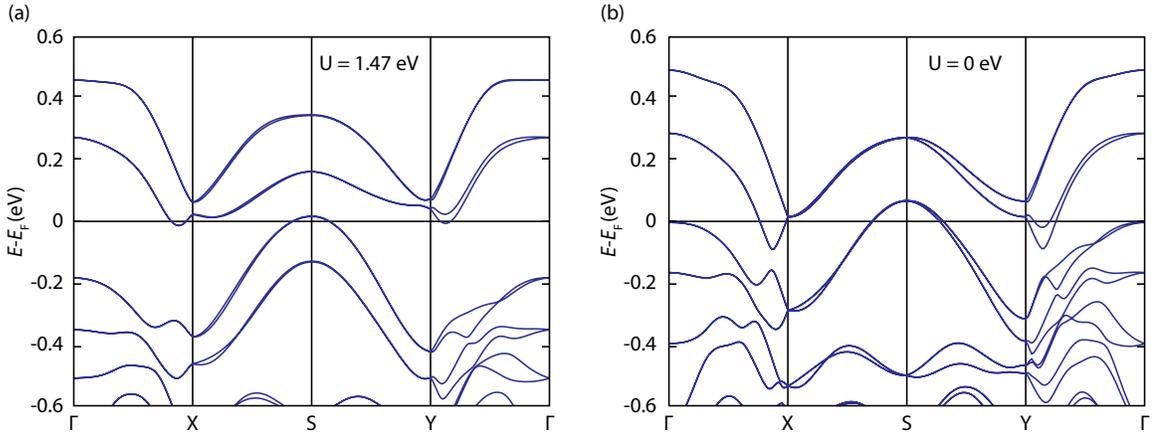

FIG. S10: SIO/STO electronic structure with the in-phase axis out-of-plane (a) with and (b) without on-site Coulomb repulsion.

volume. First, we analyze a bulk SIO unit cell (without STO) and calculate the in- and out-of-plane M−O−M bond angles with and without SOC, using experimentally found lattice constants. Next, we perform the same calculation, but we reduce the SIO unit cell volume, setting it equal to that of STO. In the calculations, the in-phase rotational axis lies along the $a$-axis and in-plane bond angles are defined as being along $a$ and $b$. The out-of-plane bond angle is along $c$. The results are shown in Table I. The M−O−M in-plane and out-of-plane bond angles are 153.7° and 154.1°, respectively for the SIO volume while they become 153.4° and 154.1°, respectively with the reduced unit cell volume. We find that changing the lattice constant does not affect the out-of-plane bond angle and only marginally reduces in-plane bond angles. In addition, inclusion of SOC does not yield a contribution larger than 0.2°.

Next we consider uniaxial strain, the results of which are reported in Fig. S11. Due to the orthorhombic unit cell of SIO, the amount of compressive strain required to lattice match SIO with STO depends on the direction along which the strain is applied. This value is at most 1.2%. Considering realistic strain values, the variations shown in Fig. S11 are less than 0.5°, which is in agreement with previous DFT data (see Fig. 3 of [18]). We conclude that neither the overall volume reduction nor application of uniaxial compressive strain seem to significantly affect the

rotation angles.

|  | STO volume | STO volume +SOC | SIO volume | SIO volume+SOC | EXP. values |
|---|---|---|---|---|---|
| in-plane | 153.4 | 153.5 | 153.7 | 153.9 | 153.5 |
| out-of-plane | 154.1 | 154.3 | 154.1 | 154.3 | 156.5 |

TABLE I: In-plane and out-of-plane M−O−M bond angles for different theoretical cases. Columns 1 and 2 show the calculated bond angles using the bulk STO unit cell volume, with and without SOC respectively. In columns 3 and 4, the bulk SIO unit cell volume is used. The last column shows experimentally found values from ref. [19]

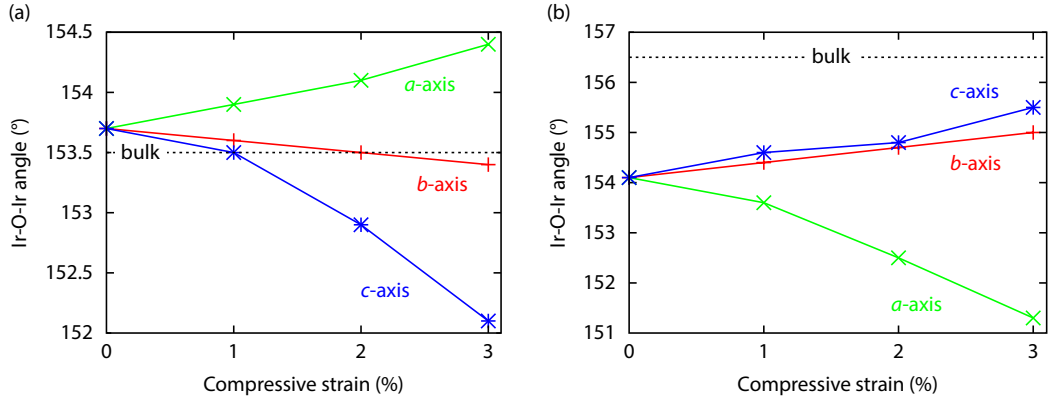

FIG. S11: (a) In-plane and (b) out-of-plane octahedral rotation angles as a function of compressive strain for bulk SIO along the $a$ (red line), $b$ (blue line) and $c$-axis (green line). The lattice constants are $a$=5.5617, $b$=5.5909 and c=7.8821 Å. The green lines represent uniaxial strain applied along the in-phase axis.

The subsequent mechanism we consider is the octahedral coupling across the STO/SIO interface. We remark that in SIO single crystals, high pressures and temperatures are required to stabilize its perovskite phase [19, 20]. STO on the other hand, is in a perovskite phase at any temperature. It is to be expected that in epitaxially stabilized STO/SIO perovskite heterostructures the SIO layer would impose distortions on the STO layers underneath. First, we analyze an $(SIO)_4/(STO)_8$ heterostructure, constraining the inner two layers of STO to be cubic (i.e. having a 180° bond angle). In bulk SIO, the two in-plane bond angles are identical. In the heterostructure however, they are found to differ slightly ($< 0.2°$). For visual clarity, the averages of the two in-plane angles are displayed. The results are shown in Fig. S12a and b, respectively. The black and red lines represent the scenario with and without constraint, respectively. In both cases, we find that the interfacial layers of the STO differ from the inner layers, meaning that Ti−O−Ti bond angles are influenced by the SIO film [21, 22]. This result matches well with the experimental observation of induced rotations in STO above the cubic-to-tetragonal transition discussed in Section II.

Lastly we consider the effect of confinement i.e., the thickness dependence of the SIO layer. To this end, we study $STO_m/SIO_n$ superstructures with a different number of SIO layers $n$, imposing periodic boundary conditions. The results are reported in Fig. S13. We find that the inner SIO layers relax to the bulk values, while the interfacial layers have slightly smaller and larger angles for the in-plane and out-of-plane rotations, respectively. For both in- and out-of-plane rotations, the angles of both the inner and the interfacial layers are found to be independent of the number of SIO layers, with the inner layers having the largest rotation angles. This suggests that interfacial coupling and not the confinement of SIO is the dominant cause of the different rotation angles. The theoretically predicted rotation angles are larger than experimentally observed. At the same time, both experiment and theory hint at the presence of a gradient of rotation angles.

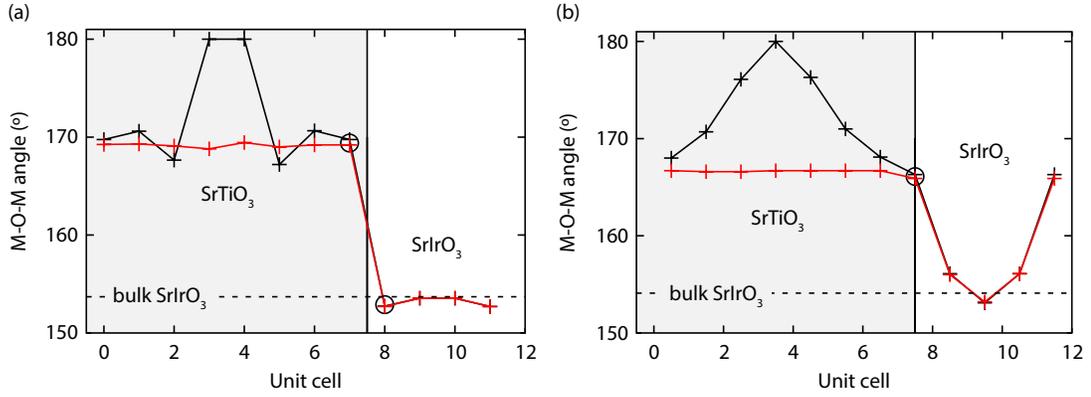

FIG. S12: (a) In-plane and (b) out-of-plane octahedral rotations along the growth axis for a $(SIO)_4/(STO)_8$ heterostructure with (black line) and without (red line) constraining inner layers of the STO to a cubic symmetry. The vertical line indicates the interface and the dashed horizontal line denotes the SIO bulk value.

We conclude that the dominant effect of the modified rotation angles of ultrathin SIO on STO with respect to the bulk values is due to the octahedral connectivity across the STO/SIO interface.

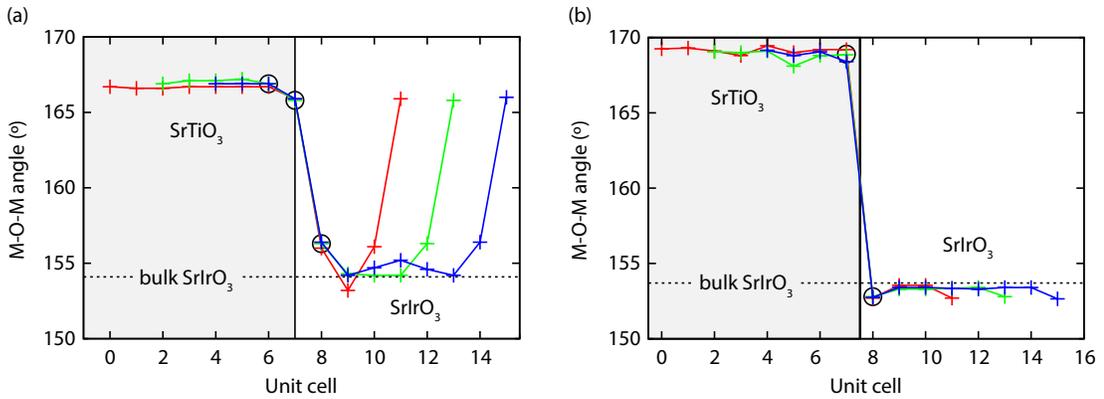

FIG. S13: (a) In-plane and (b) out-of-plane octahedral rotations along the heterostructure for the $(SIO)_4/(STO)_8$ (red line), $(SIO)_6/(STO)_6$ (green line), and $(SIO)_8/(STO)_4$ heterostructure (blue line). The vertical line indicates the interface and the dashed horizontal line denotes the SIO bulk value.



\end{document}